\documentstyle[11pt,aas2pp4]{article}

\newcommand{\gray}{{$\gamma$-ray}}
\newcommand{\grays}{{$\gamma$-rays}}

\newcommand{\mic}{$\mu$m}

\newcommand{\eqb}{\begin{eqnarray}}
\newcommand{\eqe}{\end{eqnarray}}
\newcommand{\diff}{{\rm d}}

\begin{document}


\title{Evidence for Intergalactic Absorption in the TeV Gamma-Ray 
       Spectrum of Mkn~501}

\author{Alexander K. Konopelko, John G. Kirk}

\affil{Max-Planck-Institut f\"ur Kernphysik, D-69029 
         Heidelberg, Germany}

\author{Floyd W. Stecker}

\affil{Laboratory for 
High-Energy Astrophysics, NASA/Goddard Space Flight Center, 
         Greenbelt, MD 20771}

\author{Apostolos Mastichiadis}

\affil{Department of Physics, University of Athens,
Panepistimiopolis, GR-157~83 Zografos, Greece}

\begin{abstract}

The recent {\it HEGRA} observations of the blazar Mkn~501 show 
strong curvature
in the very high energy \gray\ spectrum. Applying
the \gray\ opacity derived from an empirically based model of the 
intergalactic infrared background radiation field (IIRF), to these
observations, we find that 
the intrinsic spectrum of this source is consistent with a power-law: 
${\rm d}N_\gamma/{\rm d}E\propto E^{-\alpha}$
with $\alpha=2.00\pm0.03$
over the range $500\,$GeV -- $20\,$TeV.  
Within current synchrotron self-Compton scenarios, 
the fact that the TeV spectral energy distribution
of Mkn~501 does not vary with luminosity, 
combined with the correlated, spectrally variable 
emission in X-rays, as observed by 
the BeppoSAX and RXTE instruments,
also independently implies that the intrinsic spectrum must be 
close to $\alpha=2$. Thus, the observed curvature in the 
spectrum is most easily understood as resulting from 
intergalactic absorption. 
\end{abstract}

\section{Introduction}


Imaging air \v{C}erenkov telescopes (IACTs) are currently able to detect 
\gray\ 
photons of TeV ($10^{12}\,$eV) energy from BL~Lac objects within one hour of 
observation and to measure their spectra using a few hours of a good data (for a 
review see Weekes et al.~\cite{weekesetal97}). The objects of this class 
detected to date are, in order of increasing redshift: Mkn~421 ($z=0.031$) 
(Punch et~al.~\cite{punchetal92}, Petry et~al.~\cite{petryetal96}), Mkn~501 
($z=0.034$)
(Quinn et al.~\cite{quinnetal96}, Bradbury et al.~\cite{bradburyetal97}), 
1ES~2344+514 
($z=0.044$) (Catanese et al.~\cite{cataneseetal98}) and PKS~2155-304 ($z=0.117$)
(Chadwick et al.~\cite{chadwicketal98}). Measurements of the spectrum can be 
made  
over the energy range from $200\,$GeV to about $10\,$TeV using this technique. 
Because these photons interact with infra-red radiation to form 
electron-positron pairs, 
the signal is expected to be attenuated by absorption both within the source 
itself and 
in the intergalactic medium. Thus, it is possible to use the observations to 
study the 
intergalactic infrared radiation field (IIRF), given some general, model-dependent
constraints on the 
spectrum intrinsic to the source (Stecker, De~Jager, 
Salamon~\cite{steckeretal92}).
Determining the IIRF, in turn, allows one to model the evolution of the galaxies 
which produce it.

In this {\em Letter} we analyze recent observations of the object Mkn~501 
(Konopelko et~al.~\cite{konopelkoetal98b}) which are unique both for the quality of 
the 
spectra obtained and their energy range (up to $20\,$TeV). These data, taken 
during a
period in which the intensity of the source varied strongly, 
show a pronounced curvature in the spectrum, being 
significantly softer (steeper)
towards higher energy. 
We unfold these data using the 
upper curve for the spectral energy 
distribution of the IIRF given by Malkan \& Stecker (\cite{malkanstecker98},
henceforth MS98), with the corresponding
\gray\ opacity as calculated  
by Stecker \& De~Jager (\cite{steckerdejager98},
henceforth SD98) and find that the intrinsic 
spectrum is flat, $\diff N/\diff E \propto E^{-2}$. 
The implications of this result for both the 
absorption model and the synchro-self-Compton emission model are discussed. 

\section{Measurement of spectrum of Mkn~501}


The BL Lac object Mkn~501 exhibited strong emission in TeV $\gamma$-rays from 
March to 
October, 1997 (Protheroe et al.~\cite{protheroeetal98}). During this period the 
source 
was continuously monitored by several ground-based imaging air \v{C}erenkov 
telescopes, 
including the {\it HEGRA} stereoscopic system of 4 imaging air \v{C}erenkov telescopes 
(Aharonian et al.~\cite{aharonianetal97a}), which observed it for a 
total 
exposure time of $110\,$hours (Aharonian et al.~\cite{aharonianetal99}). The 
unprecedented statistics of about 38,000 TeV photons, combined with the good 
energy 
resolution of $\sim 20$\% over the entire energy range and with detailed studies 
of the 
detector performance (Konopelko et al.~\cite{konopelkoetal99}), allowed a 
determination 
of the spectrum in the energy range $500\,$~GeV to $24\,$~TeV (Konopelko et 
al.~\cite{konopelkoetal98b}). The Mkn~501 energy spectrum measured by the 
{\it HEGRA} collaboration extends well beyond $10\,$TeV, where 
uncertainties related to the saturation effect could in principle play a role. 
Various data consistency checks were performed in order to avoid these 
effects. In addition, simultaneous observations of the Crab Nebula 
(the standard-candle TeV source) were undertaken from 1997 September 
to 1998 March. Using similar data analysis, the Crab Nebula spectrum 
derived from the {\it HEGRA} data was found to be 
a pure power law with a differential spectrum index of 2.6 over the 
energy range $500\,$GeV -- $23\,$TeV (Konopelko et al.\ \cite{konopelkoetal98a}). 
These results are consistent with previous 
measurements by the Cangaroo group in the energy range $7$--$50\,$TeV 
(Tanimori et al.\ \cite{tanimorietal98}).


The flux of \grays, from Mkn~501, 
averaged over the entire observation period, was about 
three 
times that of the Crab Nebula ($3\,$\lq\lq Crab\rq\rq). 
Averaged over each day, the \gray\ 
rate 
showed strong variations, with a maximum of $10\,$Crab detected on 26/27 June 
1997. As 
remarked by Aharonian et al.~(\cite{aharonianetal97b}), the hardness ratio of 
the 
steepening Mkn 501 spectrum appears to be independent of the absolute flux.
The high 
\gray\ detection rate provided event statistics of a few hundreds within 1 
day's 
observations ($\sim 3-5$ hours) which suffices to evaluate the energy spectrum 
over the 
range $1$--$10\,$TeV. The analysis of the spectral shape on a daily basis did 
not reveal 
any substantial correlation between the \gray\ flux and the spectral behavior 
(Aharonian 
et al.~\cite{aharonianetal99}). This justifies the presentation of a 
time-averaged energy 
spectrum of Mkn~501 in its active state,
which is shown in Figure 1 over the 
energy range from $500\,$GeV to $24\,$TeV. The vertical error bars in this 
figure correspond 
to statistical errors. Note that the systematic errors at energies below 
$1\,$TeV appear to 
be quite large, reaching $\sim$50 \% at $500\,$GeV. 

Over the entire range, the spectrum shows a gradual softening towards higher energy. 
The $19$--$24\,$TeV energy 
bin contains a signal with a significance of $3.7\sigma$. 
However, the steep energy spectrum and
20~\% energy resolution 
do not permit one to exclude the interpretation that these \grays\ 
may have spilled over from the lower energy bins.
Therefore, the energy spectrum is consistent with the hypothesis of 
a maximum 
energy for the detected \grays\ of $\sim 18\,$TeV. The shape of the energy 
spectrum is well 
described by a power-law with an exponential cutoff. A fit of the 
data over the energy region 
where the systematic errors are small, i.e., from $1\,$TeV to $24\,$TeV, gives 
\eqb
\diff N/\diff E &=&  A E^{-\alpha}  \exp\left(-E/E_0\right)
{\rm [cm^{-2} s^{-1} TeV^{-1}]}\,. \nonumber\\
A &=& 
(9.7\, \pm 0.3\, (\textrm{stat})\, \pm 2.0\, (\textrm{syst}))\cdot 10^{-11} 
\nonumber\\
\alpha &=& {-1.9 \, \pm 0.05 \,(\textrm{stat})\, \pm 0.05 \,(\textrm{syst})} 
\nonumber\\
E_o &=& 5.7 \,\pm  1.1 (\textrm{stat}) \,\pm 0.6 \,(\textrm{syst}) {\rm TeV} 
\nonumber\\
\eqe
The logarithmic slope of the energy spectrum (``power-law index'') is $1.8$ 
in the energy 
range $1\,$--$5\,$TeV, and $3.7$ above $5\,$TeV. 

Independent measurements of the Mkn~501 TeV energy spectrum by the Whipple 
Observatory
(Samuelson et al.~\cite{samuelsonetal98}) are in very good agreement with these 
results. 
The best fit to the data presented by the Whipple group agrees precisely with a 
fit to the 
{\it HEGRA} data in the energy range $500\,$GeV -- $10\,$TeV. However the HEGRA group 
measured the 
spectrum well above $10\,$TeV where it exhibits a further steepening. 

\section{Absorption on the diffuse intergalactic infra-red background}

The formulae relevant to absorption calculations involving pair-production are 
given and 
discussed in Stecker, De Jager \& Salamon~(\cite{steckeretal92}). For \grays\ 
in the TeV 
energy range interacting at redshifts $z \ll 1$, the pair-production cross 
section 
is maximized when the soft photon energy is in the infra-red range:
\eqb
\lambda (E_{\gamma}) \simeq \lambda_{e}{E_{\gamma}\over{2m_{e}c^{2}}} &=&
2.4E_{\gamma,TeV} \; \; \mu{\rm  m} 
\eqe
where $\lambda_{e} = h/(m_{e}c)$ is the Compton wavelength of the electron. For 
a $10\,$TeV 
\gray\, this corresponds to a soft photon in the mid infra-red region of the 
spectrum, 
having a wavelength around $24\,$\mic. 
Pair-production interactions 
take place with 
photons over a range of wavelengths around the optimal value, as determined by 
the energy 
dependence of the cross section.
Stecker \& De Jager~(SD98)
have computed
the absorption coefficient of intergalactic 
space using a new, 
empirically based calculation of the spectral energy distribution (SED) of 
intergalactic low 
energy photons (MS98). 
Assuming that the IIRF is basically in 
place by a 
redshift $\sim$ 0.3, having been produced primarily at higher redshifts 
(Stecker \& De Jager~\cite{steckerdejager97},~\cite{steckerdejager98}; 
Madau~\cite{madau95}), SD98 limited their calculations to $z<0.3$.
Evolution in stellar emissivity affects the predicted IIRF and
is expected to level off 
or decrease at redshifts greater than $\sim 1.5$ (Madau \cite{madau96}). 
In this paper, we
assume that evolution continues up to $z=2$, leading to the
higher of the two IIRF used by SD98. This is more consistent with recent 
data on IR galaxy evolution, dust absorption, and the lower 
limits from IR galaxy counts (Stecker \cite{stecker99}).
To compute the absorption, we adopt the SD98 
parametric expressions for $\tau(E,z)$ for $z<0.3$, taking a 
Hubble constant of $H_o=65$ km s$^{-1}$Mpc$^{-1}$.

The unfolded {\it HEGRA} data points are also shown in Figure 1,
together with a fit to a power law energy spectrum.
We find  
\eqb
{\rm d}N_\gamma/{\rm d}E &=& 
1.32\pm0.04 \cdot 10^{-10} \times
\nonumber\\
&&(E/1\,{\rm TeV})^{-2.00\pm0.03}
\nonumber\\ 
&&{\rm photons\, cm^{-2}\, s^{-1}}
\eqe 
with $\chi^2$ = 18.6 for 15 degrees of freedom, 
giving a high chance probability of 0.2. 
In the mid-energy region $1$--$10\,$TeV, 
where the measured energy spectrum is 
very well-defined, 
the data points deviate from the fit by less then 15\%, 
which equals the estimated systematic error. 
Note that both the statistical and the 
systematic uncertainties of the spectrum measurements 
increase towards the upper end of the energy range,
where they reach 30\% and 60\%, respectively. 
Thus, the data points of the unfolded spectrum are 
consistent, within the statistical and systematic errors, 
with the simple 
power law fit of differential spectrum index 2.0. 
Our analysis shows that fitting the unfolded 
spectrum using an additional 
exponential term, result in a $\chi^2$-value of the same magnitude.   


\begin{figure}[h]
\plotone{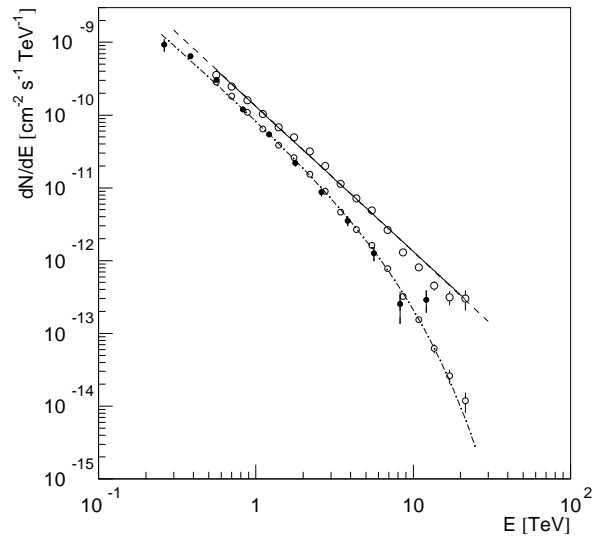}
\caption{\protect\small 
The energy spectrum of Mkn~501 as measured by the {\it HEGRA} 
IACT array 
(open circles) (Konopelko et al.~\protect\cite{konopelkoetal98b}). The combined 
power law plus 
exponent fit of the 
{\it HEGRA} 
data is shown by the dotted-dashed curve. The Mkn~501 
spectrum 
measured by the Whipple group (filled circles) is taken from 
Samuelson et al.~(\protect\cite{samuelsonetal98}). 
Also shown are the de-absorped {\it HEGRA} points (open circles)
found using the 
optical depths to absorption calculated by Stecker \& De 
Jager~(\protect\cite{steckerdejager98})
for the ``high'' intergalactic infra-red radiation field. 
The power-law fit to these data is shown by the solid line,
and has a photon index of $\protect\alpha=2.00\pm0.03$.}
\end{figure}

     
\section{Implications for the intrinsic spectrum and the inter-galactic 
absorption}

In order to understand the implications of the 
absorbed and de-absorbed spectra shown in Figure 1, 
it is necessary to adopt a model or scenario for the production of TeV 
photons in the source. Of the many suggestions in the 
literature, interest has recently centred on those in which a single 
population of relativistic 
electrons is responsible for both the TeV photons and 
for photons in the X-ray region of the spectrum, as in the 
synchrotron self-Compton (SSC) and external Compton models
(e.g., 
Bloom \& Marscher \cite{bloommarscher96};
Inoue \& Takahara \cite{inouetakahara96};
Ghisellini \& Madau \cite{ghisellinimadau96};
Dermer, Sturner \& Schlick\-eiser \cite{dermeretal97};
Mastichiadis \& Kirk~\cite{mk97} (henceforth MK97); 
Sikora et al.\ \cite{sikoraetal97};
Georganopoulos \& Marscher~\cite{gm98};
Ghisellini et al.\ \cite{ghisellinietal98}; 
Levinson \cite{levinson98}). 
These models are favored because they provide a natural way to 
understand the similar variability timescales of the X-ray and 
TeV emission.

During the period 
of the TeV observations, Mkn~501 was also observed in X-rays 
using the BeppoSAX instrument 
(Pian et al.~\cite{pianetal98}) and the Rossi X-ray Timing Explorer 
(Lamer \& Wagner~\cite{lamerwagner98}). The spectrum in this energy range 
varied strongly, with generally a very hard spectral index extending to 
much higher energies $\gtrsim100\,$keV than 
during less active epochs. There was a close temporal correlation between the 
X-ray and TeV fluxes, further 
strengthening the case for the origin of X-rays 
and TeV in a common electron population.

Extensive studies of the variability properties of the SSC model 
have been undertaken (MK97), which show that the 
mechanism most likely to be 
responsible for the variability shown by the object Mkn~421 is a change in 
the maximum energy 
(expressed as a Lorentz factor: $\gamma_{\rm max}$) 
to which electrons are accelerated. These results were also 
applied to Mkn~501 (Mastichiadis \& Kirk \cite{mk99}, 
henceforth MK99), where, 
once again, variability induced by 
a change in $\gamma_{\rm max}$ appears to give a reasonable fit
to the X-rays and to the TeV data then available.  

For such models, the most important property revealed by the {\it HEGRA}
observations discussed above is the lack of variation in the TeV spectrum,
despite the fact that the correlated variations in the X-rays 
show strong spectral variations, consistent with an 
increase in $\gamma_{\rm max}$.
Inspection of the model
light curves in the TeV range show that 
the softer the spectral slope becomes, the more sensitively it reacts to 
changes in $\gamma_{\rm max}$. This is because intrinsic 
spectra softer than a photon index of $\alpha=3$ are a direct
result of the electron
cut-off, whereas harder spectra (photon index $\alpha\approx2$) 
can be formed 
by power-law electrons scattering off
a range of target photon energies.

Thus, on the basis of the homogeneous SSC model, we find that 
intensity variations with constant spectral shape imply an intrinsic 
spectrum with a photon index of $\alpha\approx2$. 
This provides additional evidence that the spectrum of 
Mkn~501, which has a slope $\alpha\gtrsim3.7$ above $5\,$TeV, 
must be modified by the effects of inter-galactic absorption. 
Of the two IIRF considered by SD98, only the 
higher provides sufficient
absorption to account for such a strong modification. 
We note again that lower limits from galaxy counts in the 
mid-IR, as well as other observational data (Dwek et al.~\cite{dweketal98}), 
favor the high IIRF (Stecker \cite{stecker99}).

As well as constraining the intergalactic absorption, the 
observation of an intrinsic spectrum of $\alpha\approx2$ at $10\,$TeV 
requires a higher Doppler 
factor than considered by MK97 and MK99. Using the approximate
scaling laws presented in 
MK97, we can estimate that a Doppler boosting factor of 
$\delta\sim50$ suffices 
to produce $\alpha\approx2$ at $10\,$TeV, and have confirmed this by
running full simulations. It is interesting to note that such boosting 
factors, although larger 
than values measured in sources which display apparent 
superluminal motion 
(Vermeulen \& Cohen~\cite{vermeulencohen94}), seem to be indicated both by 
observations of intra-day variability (Wagner \& Witzel~\cite{wagnerwitzel95})
and of extremely rapid variations in the TeV flux of blazars (e.g., 
Gaidos et al.~\cite{gaidosetal96}). 

A completely model independent conclusion is still elusive,
and will remain so until observations 
of comparable quality on other blazars of different redshifts are available.
Nevertheless, the two independent arguments we have presented favoring an 
intrinsic emission spectrum close to $\alpha=2$ indicate that the effect of  
absorption by the intergalactic infra-red background radiation 
in the spectrum of Mkn~501 is strong, and
suggest a Doppler boosting factor for this source of $\delta\gtrsim50$.

\acknowledgments{We thank F.~Aharonian, W. Hofmann and H.J.~V\"olk for 
stimulating discussions. 
A.M. and J.G.K. acknowledge support for this collaboration 
by the European Commission under the TMR Programme, contract 
ERBFMRX-CT98-0168.

We would like to thank the anonymous referee for helpful comments.}

\end{document}